\documentclass[a4paper,11pt]{article}
\usepackage{jheppub} 
\usepackage{amssymb}                   
\usepackage{epsfig}
\usepackage{graphicx}
\usepackage{dsfont}
\usepackage[T1]{fontenc}
\def\be{\begin{equation}}
\def\ee{\end{equation}}
\def\bea{\begin{array}}
\def\eea{\end{array}}
\def\beqa{\begin{eqnarray}}
\def\eeqa{\end{eqnarray}}
\def\beqas{\begin{eqnarray*}}
\def\eeqas{\end{eqnarray*}}

\def\bp{\begin{picture}}
\def\ep{\end{picture}}
\def\bc{\begin{center}}
\def\ec{\end{center}}
\def\bfig{\begin{figure}}
\def\efig{\end{figure}}

\def\bit{\begin{itemize}}
\def\eit{\end{itemize}}
\def\nn{\nonumber}
\def\f{\frac}

\def\[{\left[}
\def\]{\right]}
\def\({\left(}
\def\){\right)}

\def\..{\left.}
\def\.{\right.}
\def\tl{\tilde}
\def\ra{\rightarrow}

\def\tm{\times}

\def\da{\dagger}

\def\al{\alpha}

\def\ep{\epsilon}

\def\Ga{\Gamma}

\def\pr{\prime}

\title{\boldmath  SUSY Breaking Constraints on Modular flavor $S_3$ Invariant $SU(5)$ GUT Model}
\author[a,1]{Xiaokang Du, Fei Wang\note{Corresponding author.}}
\affiliation[a]{School of Physics, Zhengzhou University,450000,ZhengZhou, P.R.China}
\emailAdd{feiwang@zzu.edu.cn; xkdu@gs.zzu.edu.cn}
\abstract{ Modular flavor symmetry can be used to explain the quark and lepton flavor structures.  The SUSY partners of quarks and leptons, which share the same superpotential with the quarks and leptons, will also be constrained by the modular flavor structure and show a different flavor(mixing) pattern at the GUT scale. So, in realistic modular flavor models with SUSY completion, constraints from the collider and DM constraints can also be used to constrain the possible values of the modulus parameter. In the first part of this work, we discuss the possibility that the $S_3$ modular symmetry can be preserved by the fixed points of $T^2/Z_N$ orbifold, especially from $T^2/Z_2$. To illustrate the additional constraints from collider etc on modular flavor symmetry models, we take the simplest UV SUSY-completion $S_3$ modular invariance SU(5) GUT model as an example with generalized gravity mediation SUSY breaking mechanism. We find that such constraints can indeed be useful to rule out a large portion of the modulus parameters. Our numerical results show that the UV-completed model can account for both the SM (plus neutrino) flavor structure and the collider, DM constraints. Such discussions can also be applied straightforwardly to other modular flavor symmetry models, such as $A_4$ or $S_4$ models.}
\begin{document}
\maketitle
\flushbottom

\section{Introduction}
The discovery of the Higgs boson by the ATLAS and CMS collaborations of LHC fills the last missing piece of the standard model(SM). Although SM is very successful in describing the interactions up to the electroweak scale, the origin of the mass hierarchies and mixing patterns among quarks and lepton is still unexplained. One of the most interesting solutions to such difficulties is the non-abelian discrete flavor symmetry group approach, which can lead to fairly predictive models. This approach, however, has several drawbacks.  The flavor symmetry group is usually broken
down to different subgroups by the vacuum expectation values (VEVs) of certain flavon fields. So, additional flavor symmetry breaking sector are generally needed to obtain the
desired vacuum alignment, complicated the model buildings. Moreover, large corrections from higher dimensional operators to the leading order predictions of discrete flavor symmetry models can spoil their predictability. Regarding the previous drawbacks, alternative approaches to explain the flavor structures are always welcome.

A new direction to flavor symmetry, the modular flavor symmetry, was proposed recently~\cite{Feruglio:2017spp,Criado:2018thu} to solve the flavor problem of SM. The discrete groups $S_3$,$A_4$, $S_4$ and $A_5$ are isomorphic to the finite modular groups $\Ga_N$ for $N =2,3,4,5$, respectively~\cite{deAdelhartToorop:2011re}. The Yukawa couplings are modular forms that can transform non-trivially under the finite modular group $\Ga_N$. In ref.~\cite{Criado:2018thu},  the observed mixing and mass patterns in the neutrino sector can be explained  by a flavon-free supersymmetric modular flavor symmetry model, whose superpotential can be completely determined by modular invariance. Modular form Yukawa structures transform under the finite modular
groups $S_3$~\cite{S_3}, $A_4$~\cite{Criado:2018thu,A_4}, $S_4$~\cite{S_4} and $A_5$~\cite{A_5} have been considered in the literatures.

Unfortunately, most of the economic realizations concentrate on the lepton sector and do not explain the quark flavor structure. The GUT framework, which can fit the leptons and quarks in certain GUT group multiplets, are known to be very predictive. It is therefore very interesting to survey if the modular flavor symmetry can be adopted in the GUT framework. Several works had been proposed to combine GUT with modular flavor symmetry with additional flavon fields. The interesting work of ~\cite{1906.10341}, which is based on SU(5) GUT and adopt the minimal non-abelian discrete symmetry $\Ga_2\simeq S_3$  for quark and lepton flavors, can reproduce the Cabibbo-Kobayashi-Maskawa (CKM) and Pontecorvo-Maki-Nakagawa-Sakata (PMNS) mixing matrices successfully as well as predict the leptonic CP violation phase and the effective mass of the neutrinoless double beta decay with a common  modulus parameter. Although reintroducing the flavon fields seems to lost the minimality, such models can still be very predictive.

In previous modular flavor symmetry models, supersymmetry are always adopted. The non-renormalization theorem of SUSY, which states that no new term in the superpotential will be generated in perturbation theory radiatively, can preserve the constraints of modular flavor symmetry on the superpotential in the SUSY limit. Besides, the gauge couplings unification, which can not be exact in SM, can be successfully realized in its SUSY extensions. The SUSY partners of quarks and leptons, which share the same superpotential with the quarks and leptons, can also be
constrained by the modular flavor structure and show a different flavor(mixing) pattern at the GUT scale, unlike the ordinary "SUSY breaking universality" inputs at the GUT scale adopted to evade the  flavor-changing and CP-violating constraints in the MSSM.  However, supersymmetry must be broken in realistic models. Predictions of previous mentioned modular flavor symmetry models are valid only if the SUSY breaking corrections are negligible~\cite{Criado:2018thu}. So, we need to survey the SUSY breaking effects in modular flavor symmetry models. As the soft SUSY breaking parameters are determined by SUSY breaking mechanism, it is therefore important to combine the SUSY breaking mechanism with modular flavor symmetry models and survey their low energy consequences. Depending on the way  the visible sector $'feels'$ the SUSY breaking effects in the hidden sector, the SUSY breaking mechanisms can be classified into gravity mediation~\cite{SUGRA}, gauge mediation~\cite{GMSB}, anomaly mediation~\cite{AMSB} scenarios, etc. Soft SUSY breaking parameters from economical generalized gravity mediation with non-renormalizable Kahler potential and superpotential involving
high-representation Higgs fields of various GUT group were discussed in~\cite{BNLW}. To illustrate the important constraints of SUSY breaking effects on modular GUT models, we UV-complete the modular $S_3$ SU(5) GUT model~\cite{1906.10341} as an example with the most economical generalized gravity mediation mechanism, with which the {\bf 24} representation Higgs of SU(5) fields in non-renormalizable superpotential trigger both GUT and SUSY breaking.

   This paper is organized as follows. In Sec~\ref{sec:I}, we discuss the possibility to preserve $S_3$ modular flavor symmetry by the fixed points of $T^2/Z_2$. In Sec~\ref{sec:II}, we UV-complete the modular $S_3$ SU(5) GUT model with the most economical generalized SUGRA models and obtain the resulting soft SUSY breaking parameters. In Sec~\ref{sec:III}, we give our numerical results to show the additional constraints on modular flavor symmetry from SUSY breaking. Sec~\ref{conclusion} contains our conclusions.
\section{\label{sec:I} $S_3$ modular flavor group preserved by the fixed points of $T^2/Z_N$ Orbifold}
Discrete family symmetry might come from extra dimensions. It can arise as an accidental symmetry of the orbifold fixed points or be broken by boundary conditions, see~\cite{flavor:orbifold}. Specific values for the modulus could also be obtained in extra dimensions through orbifolding~\cite{king}. We will show that $S_3$ modular group can be preserved by some of the fixed points from $T^2/Z_2$ orbifolding.

We consider the $M_4\tm T^2/Z_2$ orbifold. The extra dimensional coordinates $x_5,x_6$ are combined to be $z=x_5+ix_6$. The torus $T^2$ is defined by the complex plane $\mathbb{C}$ modulo out the lattice
\beqa
\label{torus:def}
z\sim z+\omega_1,~~z\sim z+\omega_2~,
\eeqa
with $\omega_1,\omega_2$ the lattice vectors.
Consistency of the discrete symmetry on the orbifold $T^2/Z_N$ constrains $N=2,3,4,6$~\cite{Adulpravitchai:2009id,Altarelli}.

 The inomhogeneous modular group $\bar{\Gamma}$ can be generated by
\beqa
\label{generator}
S=\(\bea{cc}~0&~1\\-1&~0\eea\)~,T=\(\bea{cc}~1&~1\\~0&~1\eea\)~.
\eeqa
which satisfy
\beqa
S^2=\mathds{1},~(ST)^3=\mathds{1}.
\eeqa

The action of $S,T$ on the basis vectors are given as
\beqa
S\(\bea{c}\omega_2\\\omega_1\eea\)=\(\bea{c}\omega_1\\-\omega_2\eea\)~,~~~
T\(\bea{c}\omega_2\\\omega_1\eea\)=\(\bea{c}\omega_1+\omega_2\\ \omega_1\eea\)~.
\eeqa
If the action of $S,T$ on the orbifold fixed points permutate the fixed points, the fixe points can form a finite representation of typical subgroup of the modular group. We should check the symmetry preserved by the orbifold fixed point.
The presentation of the finite modular group ${\Ga}_2\simeq \bar{\Ga}/\bar{\Ga}(2)$ is given as
\beqa
\Ga_2\simeq \{S,T|S^2=T^2=(ST)^3=1\}~.
\eeqa
The generator in (\ref{generator}) can also be used here. The requirement $T^2=1$ can be imposed as an constraints on the fixed points.

For $T^2/Z_2$ orbifold with basis vector $\omega_1,\omega_2$, the fix points can be found to be
\beqa
x_1=0,~x_2=\f{\omega_1}{2},~x_3=\f{\omega_2}{2},~x_4=\f{\omega_1+\omega_2}{2}.
\eeqa
Orbifold consistency conditions $z\sim -z$ for $Z_2$ would not give any constraint on the values of $\omega_2/\omega_1$.
The action of $S$ changes the fixed points as
\beqa
x_1^\pr=0,~x_2^\pr=-\f{\omega_2}{2}\sim \f{\omega_2}{2},~x_3^\pr=\f{\omega_1}{2},~x_4^\pr=\f{\omega_1-\omega_2}{2}\sim\f{\omega_1+\omega_2}{2}.
\eeqa
Such an action amounts to the permutation of the fix points
\beqa
(2~3):~(x_1,x_2,x_3,x_4)\mapsto (x_1,x_3,x_2,x_4)
\eeqa
by the $(2~3)$ element of permutation group.
The action of $T$ changes the fixed points as
\beqa
~x_1^{\pr\pr}=0,~x_2^{\pr\pr}=\f{\omega_1}{2},~x_3^{\pr\pr}=\f{\omega_1+\omega_2}{2},
~x_4^{\pr\pr}=\f{2\omega_1+\omega_2}{2}\sim\f{\omega_2}{2}.
\eeqa
Such an action amounts to the permutation of the fix points
\beqa
(x_1,x_2,x_3,x_4)\mapsto (x_1,x_2,x_4,x_3)
\eeqa
by the $(3~4)$ element of permutation group.
It is obvious that the action of $T$ is non-trivial and it satisfies $T^2=1$.  In fact, it can be checked that under $T^2$
\beqa
(T\circ T)\equiv T\cdot T\(\bea{c}\omega_2\\\omega_1\eea\)=\(\bea{c}2\omega_1+\omega_2\\ \omega_1\eea\)~,
\eeqa
the fixed points are indeed unchanged
\beqa
x_0^{\pr\pr\pr}=0,~~x_1^{\pr\pr\pr}=\f{\omega_1}{2},~~x_2^{\pr\pr\pr}=\f{\omega_2}{2}+\omega_1\sim\f{\omega_2}{2},~~
x_3^{\pr\pr\pr}=\f{\omega_1+\omega_2}{2}+\omega_1\sim\f{\omega_1+\omega_2}{2}~.
\eeqa

We also have
\beqa
(S\circ T)\equiv T\cdot S:(x_1,x_2,x_3,x_4)\mapsto(x_1,x_3,x_4,x_2)~,
\eeqa
which amounts to the $(2~3~4)$ element of permutation group.
We have the following relations for the fixed points
\beqa
S\circ S=1~,~~~T\circ T=1~,~~~(S\circ T)^3=1,
\eeqa
which is just the $S_3\simeq {\Ga}_2$ permutation group.
So, it can be seen that the fixed points of $T^2/Z_2$ can keep the $S_3$ permutation group.

For $T^2/Z_3$ orbifold under the action of $Z_3$
 \beqa
z\sim e^{i\f{2\pi m}{3}}z~,
\eeqa
for $m=0,1,2$, the orbifold consistency conditions will fix $\omega_2/\omega_1=e^{i\pi/3},e^{i 2\pi/3}, \sqrt{3}e^{i 5\pi/6}$. There are three fix points under the action of $Z_3$.
Obviously, the residual symmetry preserved by the fixed points is at most $Z_2$ because the fix point $z=0$ will not change under the action of generators on the basis vectors.
So we can arrive at the conclusion that the finite modular group $S_3\simeq \bar{\Ga}_2$ group cannot survive for fixed points set of $T^2/Z_3$.

For $T^2/Z_4$ orbifold, which fixed $\omega_2/\omega_1=e^{i\pi/2}$ by orbifold consistency conditions, can have the same type of fixed points as $T^2/Z_2$ case because all fixed points of the $T^2/Z_4$  orbifold can only appear in the second twisted sector. Similar discussions can be given and the fixed points can be compatible with the $\bar{\Ga}_2\simeq S_3$ symmetry for $\omega_2/\omega_1=e^{i\pi/2}$.

There are in total six fixed points for $T^2/Z_6$ orbifold. Fixed points from the second twisted sector are the same as that in $T^2/Z_3$ while the fixed points from the third twisted sector are the same as that in $T^2/Z_2$. It can be shown that the residual symmetry for the fixed points can at most be $S_3\times Z_2$ for $T^2/Z_6$ orbifold with $\omega_2/\omega_1=e^{i\pi/3},e^{i 2\pi/3}, \sqrt{3}e^{i 5\pi/6}$.

The transformation of superfields on orbifold for $M_4\times T^2/Z_2$ can be give by
\beqa
\phi(x,z+\omega_1)=T_5\phi(x,z)~,~~
\phi(x,z+\omega_2)=T_6\phi(x,z)~,~~
\phi(x,-z)=P_{R_\phi}\phi(x,z)~,
\eeqa
with the consistency conditions
\beqa
P_{R_\phi}^2&=&1~,~T_5 T_6=T_6 T_5~,~
T_5 P_{R_\phi}=P_{R_\phi} T_5^{-1}~, ~T_6 P_{R_\phi}=P_{R_\phi} T_6^{-1}~.
\eeqa
The $N=2$ gauge vector supermultiplets are place in the bulk with opposite parity choices for its (N=1) vector and chiral supermultiplets components to reduce the $N=1$ SUSY in 6D(amount to $N=2$ in 4D) to $N=1$ SUSY in 4D. Unlike ordinary orbifold GUT models which are adopted to solve the doublet-triplet splitting problem, the SU(5) gauge symmetry in our setting is preserved both in the bulk and in the fixed point branes so that the zero modes are still SU(5) symmetric\footnote{Our setting also allow the possibility to break $SU(5)$ into $SU(3)_c\tm SU(2)_L\tm U(1)_Y$ by non-trivial boundary condition at the fixed point $z=0$.}. The supermultiplets of three generation matter contents in SU(5) ${\bf 10}$ and $\bar{\bf 5}$ representation are placed on three different fix points of $T^2/Z_2$ (except $z=0$). The three generations, located in various fixed points, can keep the $S_3$ modular flavor symmetry.
\section{\label{sec:II}Soft SUSY breaking parameters from $S_3$ modular invariance in SU(5) GUT}
 It is possible to explain both the quark and lepton flavors by a finite modular group with a common modulus parameter $\tau$, for example in~\cite{1906.10341}. The choices of $\tau$ will be constrained by the flavor structure of the quark and lepton sectors. On the other hand, as supersymmetry is necessary to keep the form of the superpotential by the non-renormalizable theorem, the SUSY breaking effects should also be taken into account in realistic modular flavor models. Predictions of the modular flavor symmetry models are valid only if the SUSY breaking corrections are negligible~\cite{Criado:2018thu}. With proper SUSY breaking mechanism, the SUSY partners of quarks and leptons, which share the same superpotential with the quarks and leptons, could also be constrained by the modular flavor structure. Therefore, some regions of the allowed $\tau$ by flavor consideration can be favored or disfavored by additional LHC or DM constraints. We will take the simplest realistic $S_3$ modular flavor model mentioned above as an example to illustrate the main features. Besides, to generate the soft SUSY breaking spectrum in this model, the economical generalized gravity mediation SUSY breaking mechanism~\cite{BNLW} is adopted so that no additional SUSY breaking spurions are needed.

Under the modular transformation
 \beqa
\tau &\mapsto& \gamma(\tau)\equiv \f{a\tau+b}{c\tau+d}~,~~~~~~~{\rm with}~~~~ad-bc=1,
\eeqa
the superfield $\phi^{(I)}$ with modular weight $-n_{I,\phi}$ transform in a
representation $\rho^{(I)}$ of the quotient group $\Ga_N$ as follow
\beqa
\phi^{(I)} &\mapsto& \(c\tau+d\)^{-n_{I,\phi}} \rho^{(I)}(\gamma) \phi^{(I)}~.
\eeqa
So the modular invariant Kahler potential can take the following form
\beqa
K\supseteq \(-i\tau+i\overline{\tau}\)^{-n_{I,\phi}}\(\phi^{(I)\da}\phi^{(I)}\)_{\bf 1}+h \log(-i\tau+i\bar{\tau})~.
\eeqa

 Based on the modular weight choices of the various SU(5) multiplets, which are given in Table 1 with the third generation being special, the Kahler potential takes the following form\footnote{ We do not include the heavy right-handed neutrinos $N_{L;1,2,3}^c$ superfields in the Kahler potential because they will be integrated out at the low energy effective theory. They will not affect the low energy soft SUSY breaking spectrum.}
 \beqa
 K_0&\supseteq& \sum\limits_{i=1,2}\(-i\tau+i\overline{\tau}\)^{2}{\bf 10}_{i}^\da {\bf 10}_{i}+\sum\limits_{i=1,2}\(-i\tau+i\overline{\tau}\)^{2}\overline{\bf 5}_{i}^\da \overline{\bf 5}_{i}+{\bf 10}_{3}^\da {\bf 10}_{3}+\overline{\bf 5}_{3}^\da \overline{\bf 5}_{3}~,\nn\\
 &+&H_{\bf 5}^\da H_{\bf 5}+\overline{H}_{\bf \bar{5}}^\da \overline{H}_{\bf \bar{5}}
 +\Phi_{\bf 24}^\da \Phi_{\bf 24}.
 \eeqa
\begin{table}[t]
	\centering
	\begin{tabular}{|c||c|c|c|c|c|c||c|c|c||c|c|} \hline
		\rule[14pt]{0pt}{0pt}
		       & $T_{1,2}$ & $T_3$ & $F_{1,2}$ & $F_3$     & $N^c_{1,2}$ & $N^c_3$ & $H_5$ & $H_{\bar{5}}$ & $H_{24}$ & $Y_{\bf 2}^{(2)}$ & $Y^{(4)}_{\bf 1}, Y^{(4)}_{\bf 2}$ \\ \hline \hline
		\rule[14pt]{0pt}{0pt}
		SU(5)  & $10$      & $10$  & $\bar{5}$ & $\bar{5}$ & 1           & 1       & $5$   & $\bar{5}$     & $24$     & 1                 & 1                                  \\
		$S_3$  & 2         & $1'$  & 2         & $1'$      & 1           & $1'$    & 1     & 1             & 1        & 2                 & 1,\ 2                              \\
		weight & $-2$      & $0$   & $-2$      & $0$       & 0           & 0       & $0$   & $0$           & $0$      & $2$               & $4$                                \\ \hline
	\end{tabular}
	\caption{ In order to compare the constraints from SUSY and modular flavor symmetry, we choose the same charge assignments of SU(5), $S_3$ and modular weight for superfields (and modular forms) as that from ref.(\cite{1906.10341}). The subscript $i$ of $F_i$ and $T_i$ denotes the $i$-th family.}
	\label{tb:fields}
\end{table}
It is obvious from the Kahler potential that the kinetic terms are not canonical. So we need to rescale the fields by a wavefunction normalization factor $Z_\phi$ to obtain the proper forms of the physical spectrums. From the Kahler potential, we can obtain the normalization factor
 \beqa
\label{normalization}
 Z_{Q_L;1,2}&=&Z_{U_L^c;1,2}=Z_{E_L^c;1,2}=Z_{D_L^c;1,2}=Z_{L_L;1,2}={4 (Im~ \tau)^2}~,~ \nn\\ Z_{Q_L;3}&=&Z_{U_L^c;3}=Z_{E_L^c;3}=Z_{D_L^c;3}=Z_{L_L;3}=1,\nn\\
 Z_{H_u}&=&Z_{H_d}=1~.
 \eeqa
The value of $v_{24}/\Lambda$ was set to be $0.3$ in \cite{1906.10341} to generate additional contributions to the flavor structure. Here we neglect the subleading terms proportional to $v_{24}^2/\Lambda^2\sim 0.1$ in the normalization factors.

To adopt the generalized gravity mediation scenario, the {\bf 24} representation Higgs of SU(5) is assumed to trigger both the SU(5) gauge symmetry breaking and the SUSY breaking. The lowest and F-component VEVs of the adjoint Higgs of SU(5) can take the following form
\beqa
\langle \Phi_{\bf 24}\rangle=\(v_{\bf 24}+\theta^a F_{\bf 24}\)\sqrt{\f{3}{5}}\left(
\bea{ccccc}\f{1}{3}&&&&\\&\f{1}{3}&&&\\&&\f{1}{3}&&\\&&&-\f{1}{2}&\\&&&&-\f{1}{2}\eea\right).
\eeqa
 Modular invariance also constrain to some extent the non-renormalizable terms involving the {\bf 24} representation Higgs in the Kahler potential
 \beqa
\label{sfermion}
 K\supseteq K_0\(1+{c_0}\(-i\tau+i\overline{\tau}\)^{-n_{\bf 24}}\f{Tr(\Phi_{\bf 24}^\da\Phi_{\bf 24})}{\Lambda^2}\)~,
 \eeqa
although additional terms involving the $Y_i^{(2)},Y_i^{(4)}$ etc can still be consistent with the modular flavor symmetry of the model. As noted in~\cite{Chen:2019ewa}, the presence of such additional terms will reduce the predictive power of these modular group constructions. So we simply assume that proper mechanism will forbidden other terms to appear in the non-renormalizable Kahler potential (\ref{sfermion}).
We should note that the normalization factors will eventually be canceled out in the final expressions although the kinetic terms are not canonical.  So the (universal) soft SUSY breaking masses for the sfermions (at the GUT scale) are given as
 \beqa
 m^2_{\tl{\phi}}= \f{c_0}{2}\f{\left|F_{\bf 24}\right|^2}{\Lambda^2}~.
 \eeqa

To introduce the higher-dimensional operators involving {\bf 24} representation Higgs fields, we need to know the decompositions of the tensor products of various SU(5) representations for
the Yukawa coupling terms~\cite{BNLW}
\beqa
{\bf 10}\otimes{\bf 10}\otimes{\bf
5}&=&({\bf \bar{5}} \oplus
{\bf \overline{45}}\oplus{\bf \overline{50}} )\otimes {\bf 5} \nonumber
\\
&=& ({\bf 1} \oplus {\bf 24} ) \oplus
({\bf 24}\oplus {\bf 75}\oplus{\bf 126}) \oplus
({\bf 75}\oplus {\bf 175'}) ~,~ \\
{\bf 10}\otimes{\bf \bar{5}}\otimes {\bf \bar{5}}&=&{\bf
10}\otimes({\bf \overline{10}\oplus \overline{15}})=({\bf 1 \oplus
24\oplus
75})\oplus({\bf 24 \oplus \overline{126} })~.
\eeqa
We can see that there are two types of contraction of Yukawa with adjoint representation Higgs $\Phi_{\bf 24}$ for both Yukawa couplings. The most general discussions on soft SUSY parameters related to the introduction of higher-dimensional operators can be seen in our previous works. Although not shown explicitly, the contraction rules within ~\cite{1906.10341} adopt the following contractions
\beqa
\label{contraction}
W\supseteq {\bf 10}_{ij}{\bf 10}_{mn} {\bf 5}_{k} \f{{\bf 24}^k_{l}}{\Lambda}\epsilon^{ijmnl}
+{\bf 10}_{ij}\overline{\bf 5}^{k} \overline{\bf 5}_{H}^j \f{{\bf 24}^i_{k}}{\Lambda}~.
\eeqa
Modular invariance also constrain stringently the form of the superpotential. The modular weight choices and the modular flavor group representation choices can fix the form of the superpotential except the non-renormalizable terms involving $\Phi_{\bf 24}$. To compare the constraints from SUSY and contraints from the SM flavor structures, we adopt the same form of non-renormalizable terms involving $\Phi_{\bf 24}$ as that in~\cite{1906.10341}.

 The $A_u$ trilinear couplings can be obtained by substituting the F-component VEV of
$\Phi_{\bf 24}$ into the superpotential involving $10-10-5_H$ couplings of SU(5)\footnote{The mass matrix for up-type quarks can take an asymmetric form with additional {\bf 45} representation Higgs fields.}
 \beqa
 W_1\supseteq \(\al_1^\pr Y_1^{(4)}+\al_2^\pr Y_2^{(4)}\)T_{1,2} T_{1,2} H_5 k_1^\pr\f{\Phi_{\bf 24}}{\Lambda}
 +  \beta_1^\pr Y_2^{(2)}T_{1,2} T_3 H_5 k_2^\pr\f{\Phi_{\bf 24}}{\Lambda}
 +\gamma^\pr T_{3} T_3 H_5 k_4^\pr\f{\Phi_{\bf 24}}{\Lambda}~,\nn\\
 \eeqa
with the production rule of {\bf 2} representation of $S_3$ given as
\beqa
\(\bea{cc}x_1\\x_2\eea\)\otimes\(\bea{cc}y_1\\y_2\eea\)=(x_1y_1+x_2y_2)_{\bf 1}\oplus(x_1y_2-x_2y_1)_{\bf 1^\pr}\oplus \(\bea{cc}x_1y_1-x_2y_2\\-x_1y_2-x_2y_1\eea\)_{\bf 2}~,
\eeqa
and the expressions of $Y_1,Y_2$ etc given in the appendix.  The trilinear couplings are therefore given by
 \beqa\scriptsize
 (\tl{A}_u)_{ij}=-{2}\sqrt{\f{3}{5}}\f{F_{\bf 24}}{\Lambda}\left(\bea{ccc}\[\al_1^\pr(Y_1^2+Y_2^2)+\al_2^\pr(Y_1^2-Y_2^2)\]k_1^\pr& ~~2\al_2^\pr Y_1 Y_2k_1^\pr~~&~~  -k_2^\pr\beta_1^\pr Y_2~~\\2\al_2^\pr Y_1 Y_2k_1^\pr &\[\al_1^\pr(Y_1^2+Y_2^2)-\al_2^\pr(Y_1^2-Y_2^2)\]k_1^\pr& k_2^\pr\beta_1^\pr Y_1 \\ -k_2^\pr\beta_1^\pr Y_2 & k_2^\pr\beta_1^\pr Y_1& k_4^\pr\gamma^\pr\eea\right)\nn\\\normalsize
 \eeqa

The $\tl{A}_d$ trilinear couplings can be obtained from the $10-\bar{5}-\bar{5}_H$ coupling
\beqa
W_2&\supseteq &\(\al_1 Y_1^{(4)}+\al_2 Y_2^{(4)}\)T_{1,2} F_{1,2} {H}_{\bar 5} k_1 \f{\Phi_{\bf 24}}{\Lambda}
 +  \beta_1 Y_2^{(2)}T_{1,2} F_3 {H}_{\bar 5} k_2 \f{\Phi_{\bf 24}}{\Lambda}+
 \beta_2 Y_2^{(2)}T_3 F_{1,2} {H}_{\bar 5} k_3 \f{\Phi_{\bf 24}}{\Lambda}\nn\\
 &+&\gamma T_{3} F_3 {H}_{\bar 5} k_4 \f{\Phi_{\bf 24}}{\Lambda}~,
\eeqa
after substituting the F-term VEV of $\Phi_{\bf 24}$ with the contraction given in (\ref{contraction}). It reads
\beqa\scriptsize
(\tl{A}_d)_{ij}=\f{1}{3}\sqrt{\f{3}{5}}\f{F_{\bf 24}}{\Lambda}\left(\bea{ccc}\[\al_1(Y_1^2+Y_2^2)+\al_2(Y_1^2-Y_2^2)\]k_1 & ~~2\al_2 Y_1 Y_2k_1~~&~~ -k_2 \beta_1 Y_2 ~~\\2\al_2 Y_1 Y_2k_1 &\[\al_1(Y_1^2+Y_2^2)-\al_2(Y_1^2-Y_2^2)\]k_1 & k_2 \beta_1 Y_1 \\ -k_3\beta_2  Y_2 &k_3\beta_2  Y_1 &  \gamma k_4\eea\right).\nn\\\normalsize
 \eeqa
The trilinear couplings for the lepton Yukawa couplings can be similarly obtained to be
 \beqa
 (\tl{A}_e)^T_{ij}=-\f{3}{2} (\tl{A}_d)_{ij}~.
 \eeqa

Due to the fact that the superfields in the superpotential are not canonically normalized, proper wavefunction factors should be divided from the previous expressions of trilinear terms to get the physical ones. So, the genuine trilinear soft terms are given as
\beqa
{A}^{u,d,e}_{ijk}=\f{\tl{A}^{u,d,e}_{ijk}}{\sqrt{Z_iZ_jZ_k}},
\eeqa
with $Z_\phi$ given in (\ref{normalization}).

 Unlike the KKLT realization, in which gauge fields origin from the D3/D7 brane, the gauge fields here need not transform under modular transformations.
The gaugino masses are generated by non-renormalizable term involving the $\Phi_{\bf 24}$ Higgs.
 \beqa
{\cal L}\supseteq \int d^2\theta \f{1}{4}W^a W^a-\f{c_1}{4\Lambda} W_\mu^a \Phi^b_a W_{\mu}^b.
 \eeqa
 After the {\bf 24} representation Higgs $\Phi$ acquires VEV, the gaugino masses are given as
 \beqa
 M_i=-\f{c_1}{Z_i} a_i\f{F_{\bf 24}}{12\Lambda}\sqrt{\f{3}{5}}~,
 \eeqa
 with $a_i=(~1,~3,-2)$  for $U(1)_Y,SU(2)_L,SU(3)_c$ and the wavefunction normalization factor
 \beqa
 Z_i=1-a_i\f{c_1}{12\Lambda}\sqrt{\f{3}{5}}v_{24}\approx 1.
 \eeqa
\section{\label{sec:III}Numerical Results}
This minimal UV-completion $S_3$ modular flavor SUSY GUT model has three additional SUSY breaking related free parameters $c_0,c_1,{F_{\bf 24}}/{\Lambda}$ in addition to the modulus parameter $\tau$ and other parameters chosen in the $S_3$ modular SU(5) GUT model~\cite{1906.10341} to generate the flavor structure for quarks and leptons.
As noted previously, constrains on the superpotential by the modular flavor structure will lead to "non-universality" of the SUSY breaking parameters for the SUSY partners of quarks and leptons at the input scale, which share the same superpotential with the quarks and the leptons. Therefore, a different flavor pattern at the EW scale (in contrast to "universality" soft SUSY breaking inputs) will be generated and low energy flavor constraints etc will impose non-trivial constraints on the input parameters. So, it is interesting to survey if this SUSY completion model can survive the low energy (SUSY) flavor, collider as well as the DM constraints.

The package SPheno4.0.4\cite{SPheno} is used to scan the parameters spaces of
\beqa
\tau,~c_0,~c_1,~\f{F_{\bf 24}}{\Lambda},
\eeqa
 with the model file generated by SARAH4.14.3 \cite{SARAH}. The DM relic density and other collider constrains are obtained with the package micrOMEGA5.2.4\cite{micrOMEGA}. We use the package HiggsBounds5.3.2\cite{HiggsBounds} to check whether or not the predictions of the survived points can pass the exclusion bounds obtained from the Higgs searches by the colliders.

We impose the following constraints in our numerical studies:
\bit
\item (I) The conservative lower bounds from current LHC constraints on SUSY particles\cite{atlas:gluino,atlas:stop}, the Z-boson invisible decay and the LEP search for sparticles\cite{LEP} as well as Higgs mass
\beqa
m_{\tl{t}_1} \gtrsim 1.2 {\rm TeV}, ~m_{\tl{g}} \gtrsim 2.2{\rm TeV},
 m_{\tl{b}_1} \gtrsim 0.95 {\rm TeV},~ m_{\tl{u}_{1}},~m_{\tl{d}_{1}} \gtrsim 1.37 {\rm TeV},\nn\\
m_h\in 125\pm  2 {\rm GeV},~ m_{\tl{\chi}^\pm}> 103.5 {\rm GeV}~,\Gamma(Z\ra \tl{\chi}_0\tl{\chi}_0)<1.71~{\rm MeV}~.
\eeqa
\item  (II) Constraints from various flavor constraints, such as the rare B mesons decay \cite{B-physics} and lepton flavor violation(LFV) bounds~\cite{mu-egamma,tau-emugamma,mu-3e,tau-3mu3e}:
  \beqa
  && 0.85\tm 10^{-4} < Br(B^+\ra \tau^+\nu) < 2.89\tm 10^{-4}~,\nn\\
  &&1.7\tm 10^{-9} < Br(B_s\ra \mu^+ \mu^-) < 4.5\tm 10^{-9}~,\nn\\
  && 2.99\tm 10^{-4} < Br(B_S\ra X_s \gamma) < 3.87\tm 10^{-4}~,\nn\\
  && Br(\mu \rightarrow e  \gamma) < 4.2 \times 10^{-13}~,~~
   Br(\tau \rightarrow e \gamma) < 3.3 \times 10^{-8}~, \nn\\
&& Br(\tau \rightarrow \mu \gamma) < 4.4 \times 10^{-8}~,~~
   Br(\mu \rightarrow 3e) < 1.0 \times 10^{-12}~, \nn\\
&& Br(\tau \rightarrow 3e) < 2.7 \times 10^{-8}~,~~
   Br(\tau \rightarrow 3\mu) < 2.1 \times 10^{-8}~.
   \eeqa
\item (III) The relic density of dark matter(DM) should satisfy the Planck data $\Omega_{DM} = 0.1199\pm 0.0027$ \cite{Planck} in combination with the WMAP data \cite{WMAP}(with a $10\%$ theoretical uncertainty).
\eit

We have the following discussions related to our numerical results:
\bit

\begin{figure}
\centering
\includegraphics[width=.49\textwidth]{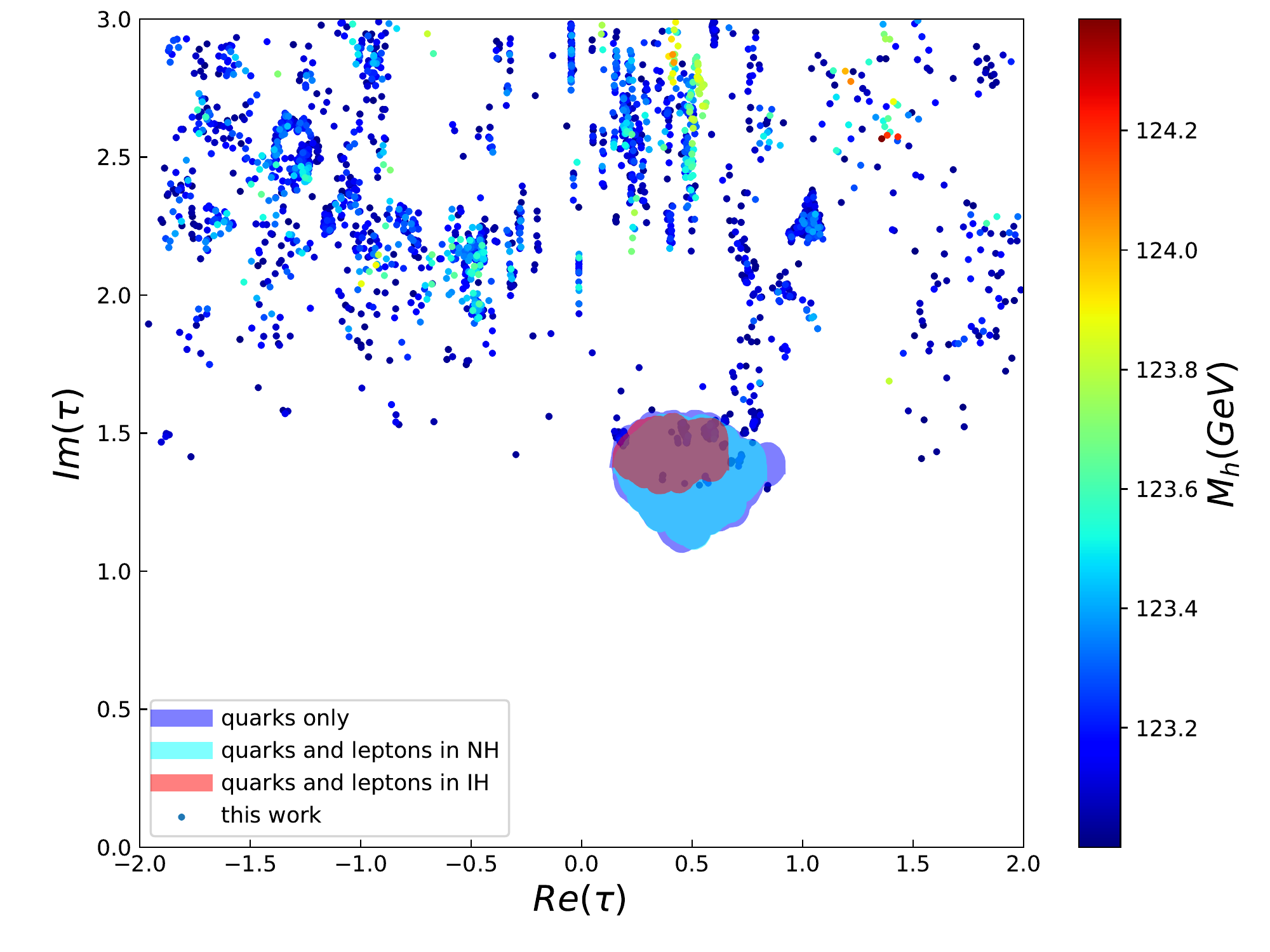}
\includegraphics[width=.49\textwidth]{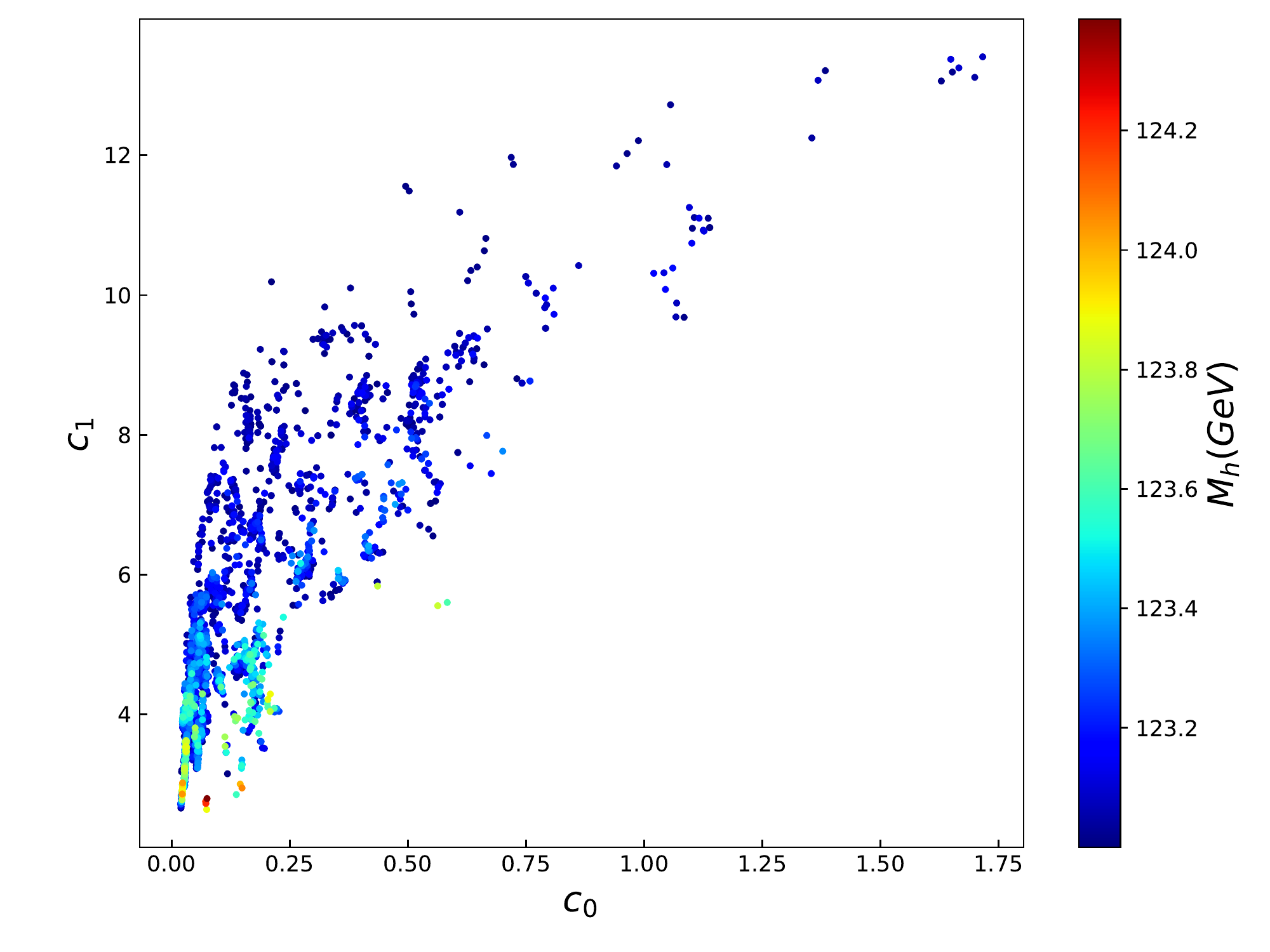}
\caption{In the left panel, we show the allowed range of $Im(\tau)$ versus $Re(\tau)$ that can satisfy the SM flavor structure, the collider and DM constraints. The best-fit region of quark flavor structure in \cite{1906.10341} is labeled with blue color and the best-fit regions for both quark and lepton flavor structures with neutrino mass pattern either in IH (NH) is labeled with red color (cyan color), respectively. In the right panel, we show the survived range of $c_0,c_1$ parameter with the corresponding Higgs masses.}
\label{fig1}
\end{figure}

\item It can be seen from the left-panel of fig.\ref{fig1} that the constraints from (I) to (III) can rule out much of the allowed parameter space of $\Im(\tau)$ versus $\Re(\tau)$, especially a large portion of the best-fit region of the SM(plus neutrino) flavor structure in \cite{1906.10341}. On the other hand, it is interesting to note that the survived region of $\Im(\tau)$ versus $\Re(\tau)$ can still have a overlap with such best-fit region, which means that the UV-completed model can account for both the SM (plus neutrino) flavor structure and the collider, DM constraints.

A tiny fraction of allowed parameter space can lie within the best-fit region for quark flavor structure, which is label with blue color.
  The best-fit regions for both quarks and leptons flavor structures with neutrino mass pattern either in invert hierarchy(IH) (label with red color) or normal hierarchy(NH) (label with cyan color) can also be consistent with the collider and DM constraints. As the best-fit region for IH case has large overlap with the best-fit region for NH case, constraints from the the collider and DM still not show any preference between the two cases. If more experimental data for the SM (plus neutrino) flavor structure can be released in the future to separate the best-fit region for NH case and IH case, SUSY breaking consideration can possibly show which type of neutrino mass hierarchy is more preferable.
\item The allowed range of $c_0 ,c_1$ is shown in the right-panel of fig.\ref{fig1} with the corresponding Higgs mass. It is obvious from the figure that the discovered 125 GeV Higgs can also be accommodated in this SUSY UV-completion framework. From the right panel of fig.\ref{fig1}, we can see that the predicted Higgs mass is always below 125 GeV. Taking into account the theoretical uncertainty in calculating the Higgs mass, which is about $0.5\sim 2$ GeV, the best-fit region of $\Im(\tau)$ versus $\Re(\tau)$ for flavor structures can also marginally accommodate the 125 GeV Higgs mass.

\begin{figure}
\centering
\includegraphics[width=.49\textwidth]{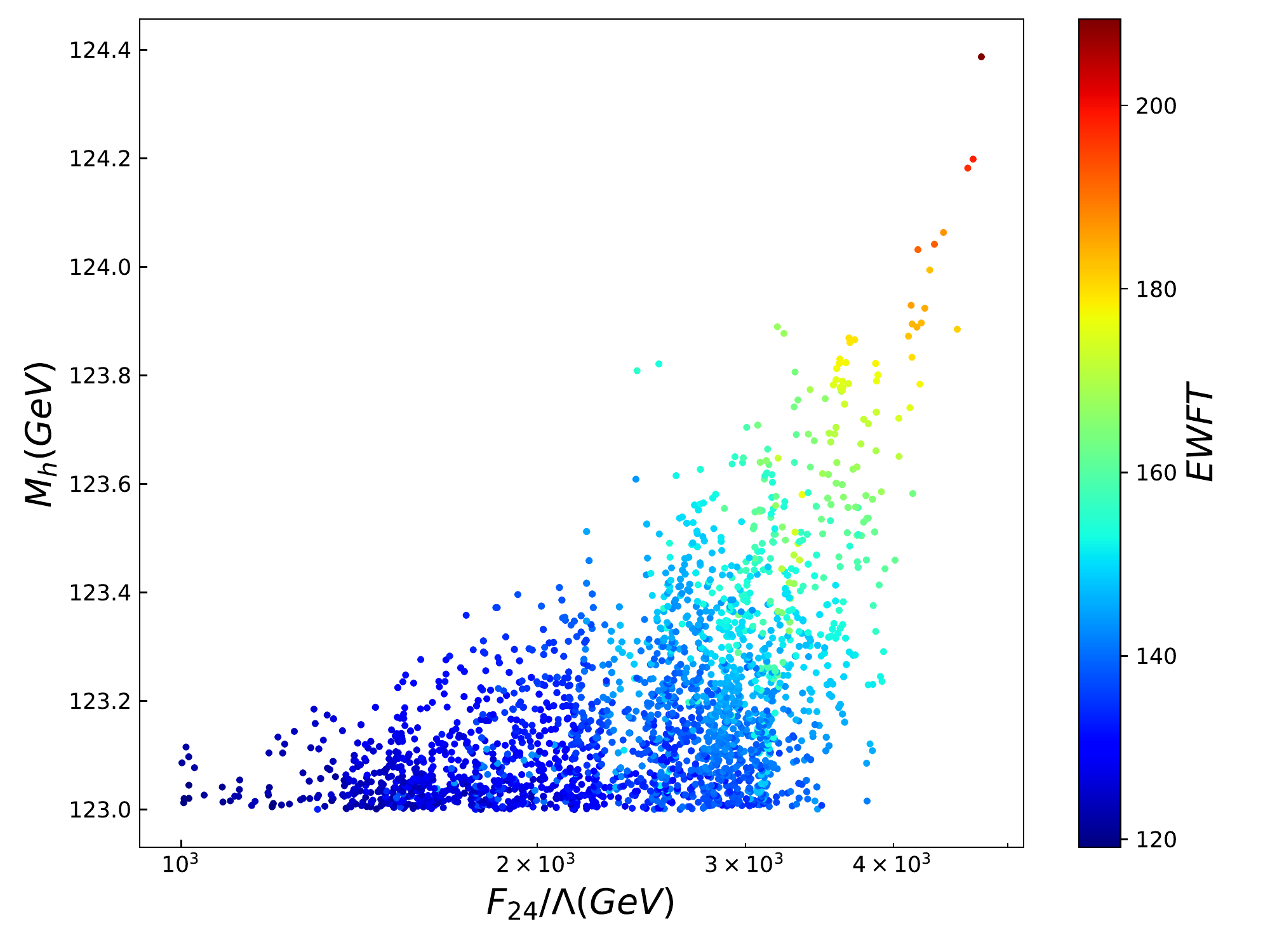}
\includegraphics[width=.49\textwidth]{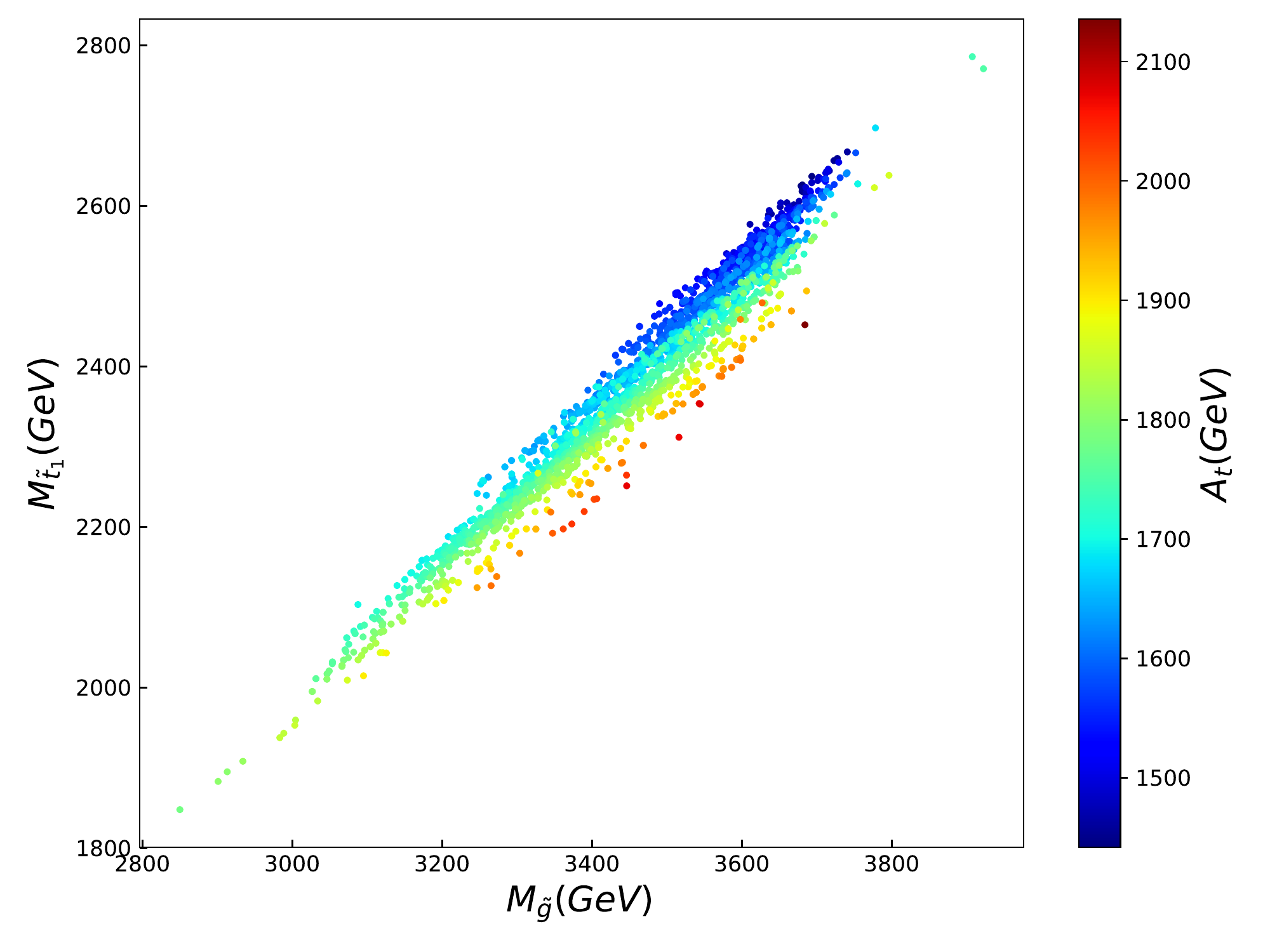}\\
\includegraphics[width=.49\textwidth]{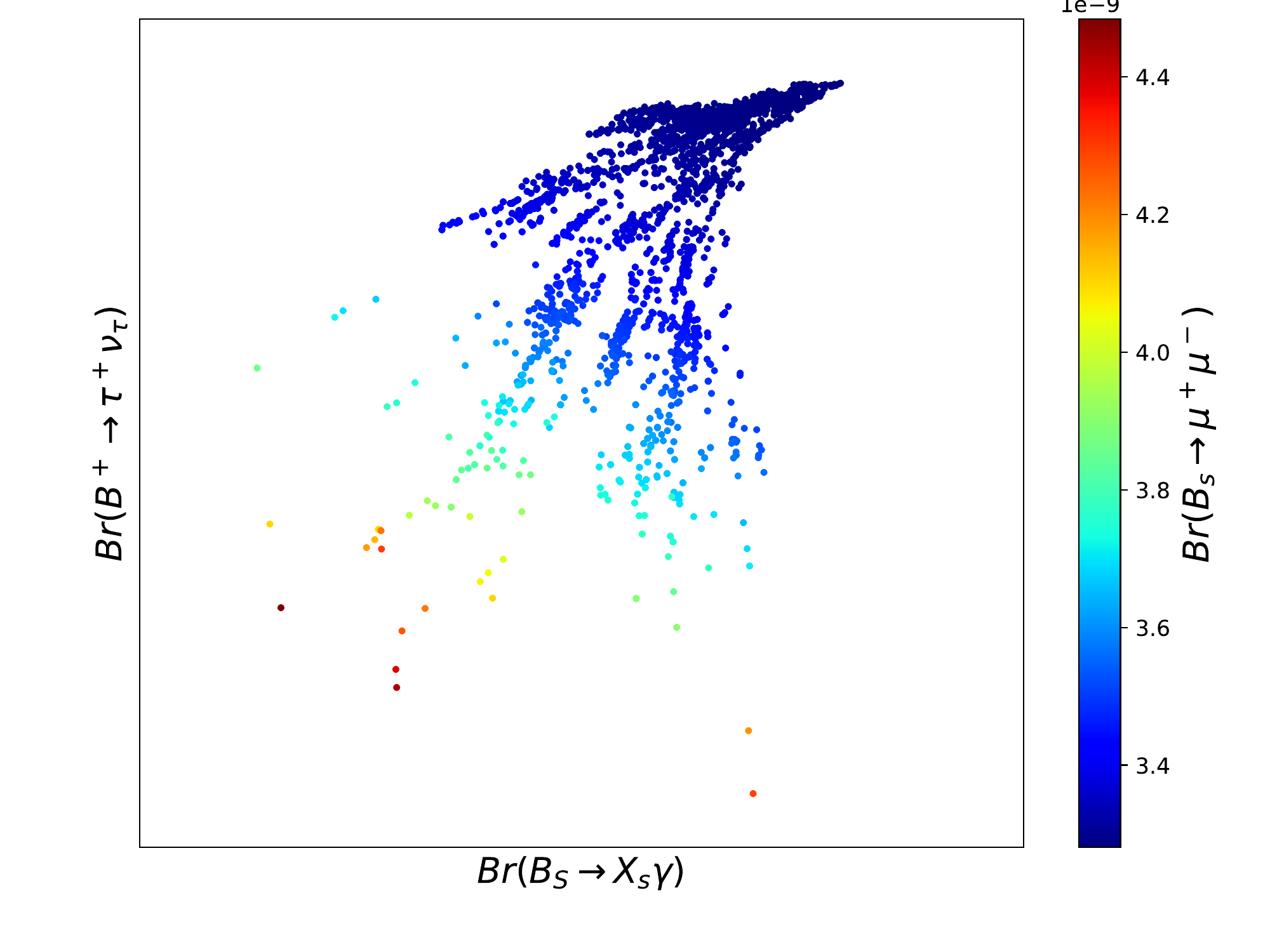}
\includegraphics[width=.49\textwidth]{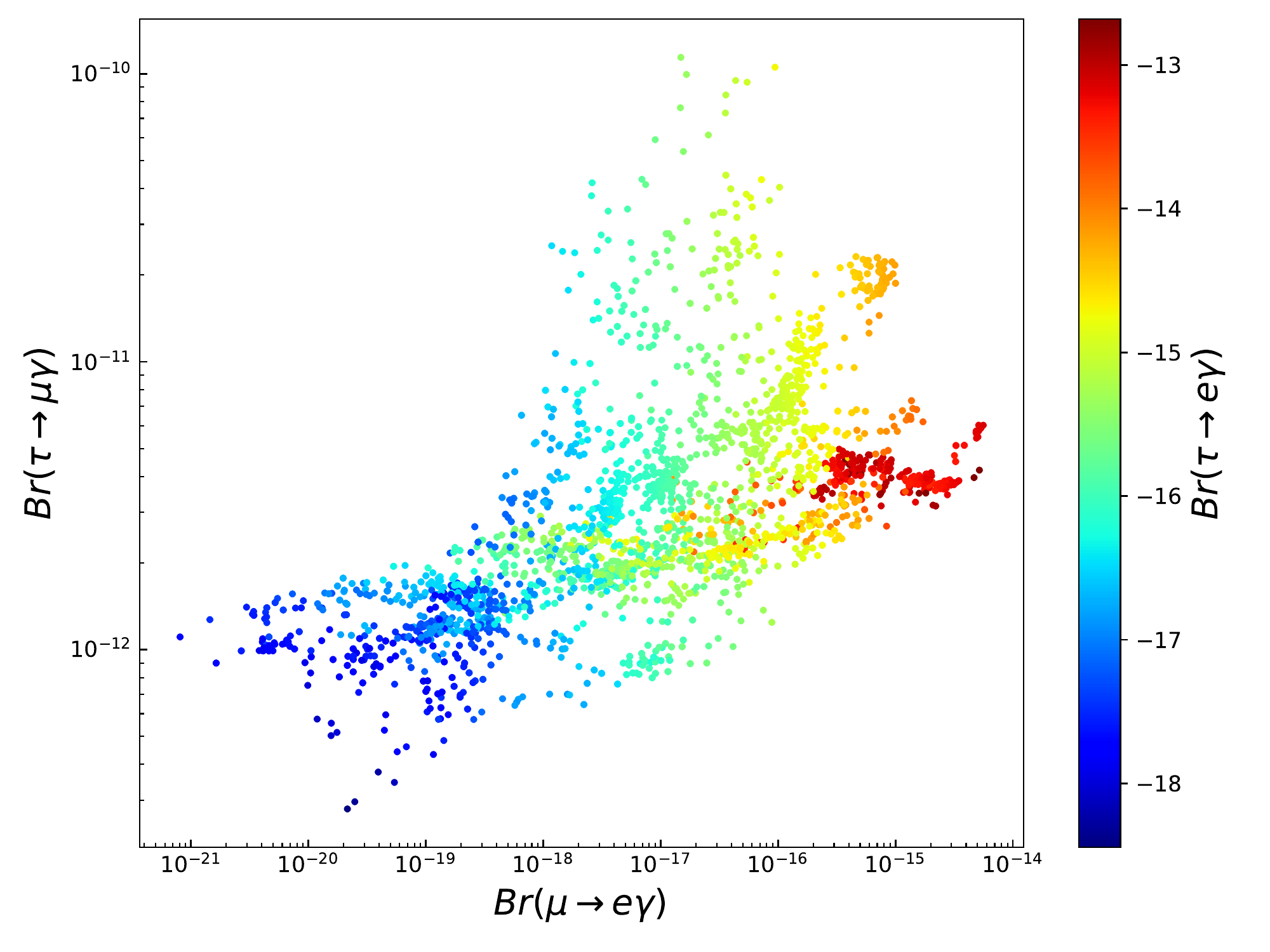}
\caption{The survived regions for the mass parameter $F_{\bf 24}/\Lambda$ versus the Higgs mass are shown in the upper left panel. Survived regions for gluino masses versus the lightest stop masses are shown in the upper right panel. Branch ratios of various new physics sensitive B meson and LFV  decay modes are shown in the lower left and right panel, respectively.}
\label{fig2}
\end{figure}
\item The survived region for the mass parameter $F_{\bf 24}/\Lambda$, which is the only new mass parameter that determines the soft SUSY breaking scale at the GUT scale, is constrained to lie almost between 1.0 TeV to 4.8 TeV (see the upper-left panel of fig.\ref{fig2}). The values of $F_{\bf 24}/\Lambda$ and $c_0,c_1$ can determine the whole soft SUSY spectrum. Our numerical results indicate that the gluino masses are predicted to lie between 2.85 TeV to 3.92 TeV. The lighter stop, $\tl{t}_1$, are predicted to lie between 1.85 TeV to 2.78 TeV. Both of them can be possibly tested on the near future upgraded LHC. Because new flavor mixing and CP violating structures will appear in the squark and slepton sector (in contrast to the "universality " soft SUSY breaking inputs), some of the low energy flavor constraints will become important. The branch ratios of typical SUSY-sensitive B meson and LFV decay modes are  shown in the lower panels of fig.\ref{fig2}, which again can be checked in the near-future collider experiments.

\begin{figure}
\centering
\includegraphics[width=.49\textwidth]{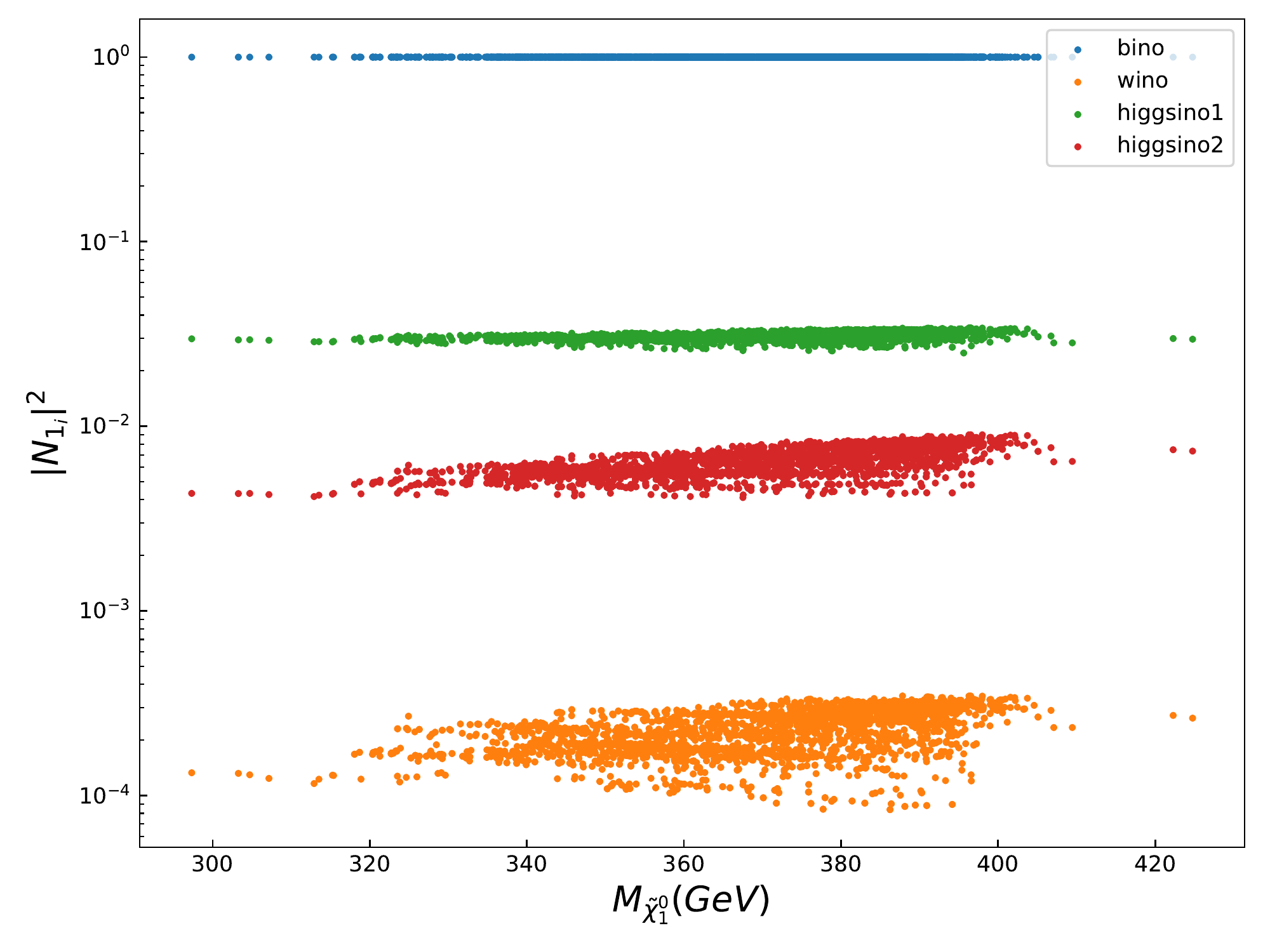}
\includegraphics[width=.49\textwidth]{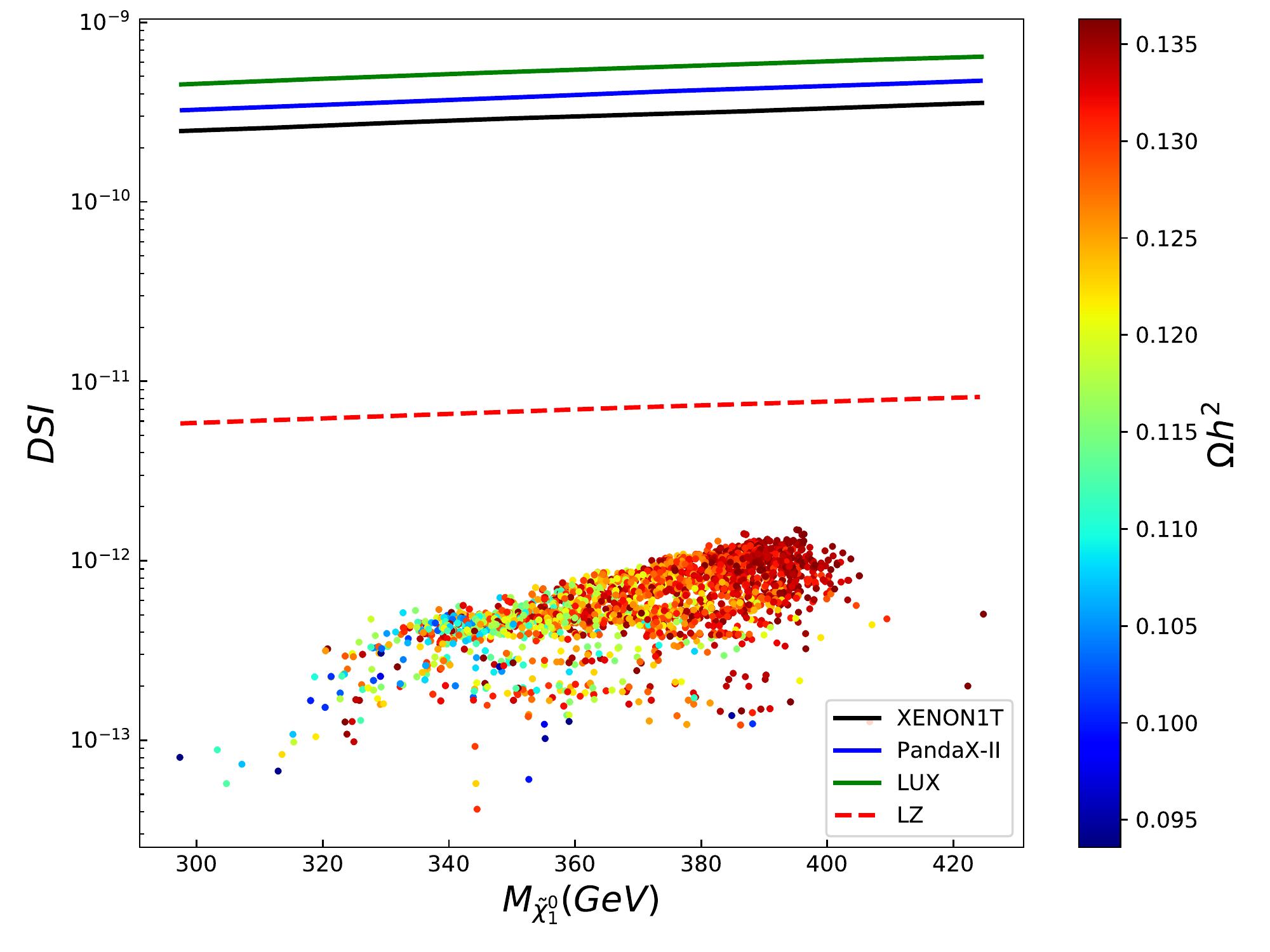}
\caption{In the left panel, relative components of the LSP are shown. In the right panel, the SI DM-nucleon scattering cross section for DM direct detection is shown. Exclusion bounds of various direct detection experiments, such as XENON1T, PANDAX and LZ, are also given. }
\label{fig3}
\end{figure}
\item   From our numerical results, it can also be seen that correct DM relic abundance can be obtained and the DM candidate is always the lightest neutralino, which is mostly bino-like(see the left-panel of fig.\ref{fig3}). As pure bino-like DM can not annihilate into W or Z gauge bosons, they can only annihilate into standard model fermions via t-channel sfermions exchange. So, unless bino is lighter than 300 GeV, pure bino-like DM will always lead to overabundance of DM relic density. On the other hand, coannihilation with stau can reduce the DM abundance efficiently. We checked that the bino-stau coannihilation effect is very crucial to obtain correct DM relic density. The spin-independent(SI) DM-nucleon cross section for DM direct detection is shown in the right-panel of fig.\ref{fig3}.  We can see that most of the allowed parameter space will survived the current DM direct detection bounds, such as LUX\cite{LUX},XENON1T\cite{XENON1T} and PANDAX\cite{PANDAX}. Besdies, they can even survive the near-future LUX-ZEPLIN\cite{LZ} experimental bounds. So, no DM signal is anticipated to be found in the near-future DM direct detection experiments.
\eit

\section{\label{conclusion}Conclusions}

Modular flavor symmetry can be used to explain the quark and lepton flavor structure.  The SUSY partners of quarks and leptons, which share the same superpotential with the quarks and leptons, will also be constrained by the modular flavor structure and show a different flavor(mixing) pattern at the GUT scale. So, in realistic modular flavor models with SUSY completion, constraints from the collider and DM constraints can also be used to constrain the possible values of the modulus parameter. In the first part of this work, we discuss the possibility that the $S_3$ modular symmetry can be preserved by the fixed points of $T^2/Z_N$ orbifold,  especially from $T^2/Z_2$. To illustrate the additional constraints from collider etc on modular flavor symmetry models, we take the simplest UV SUSY-completion $S_3$ modular invariance SU(5) GUT model as an example with generalized gravity mediation SUSY breaking mechanism. We find that such constraints can indeed be useful to rule out a large portion of the modulus parameters. Our numerical results show that the UV-completed model can account for both the SM (plus neutrino) flavor structure and the collider, DM constraints.
Such discussions can also be applied straightforwardly to other modular flavor symmetry models, such as $A_4$ or $S_4$ models.

\begin{acknowledgments}
 This work was supported by the Natural Science Foundation of China under grant numbers
12075213,11675147,11775012 and the Key Project by the Education Department of Henan Province under grant number 21A140025.

\end{acknowledgments}
\appendix
\section*{Appendix}
\section{Expressions of the modular forms for $S_3$ modular group}
For completeness, we collect the expressions of modular forms with different modular weight from \cite{1906.10341}. The modular forms of weight 2 corresponding to the $S_3$ doublet are written as
\begin{eqnarray}
	\label{eq:Y-S3}
	Y_1(\tau) &=& \frac{i}{4\pi}\left( \frac{\eta'(\tau/2)}{\eta(\tau/2)}  +\frac{\eta'((\tau +1)/2)}{\eta((\tau+1)/2)}
	- \frac{8\eta'(2\tau)}{\eta(2\tau)}  \right), \nonumber \\
	Y_2(\tau) &=& \frac{\sqrt{3}i}{4\pi}\left( \frac{\eta'(\tau/2)}{\eta(\tau/2)}  -\frac{\eta'((\tau +1)/2)}{\eta((\tau+1)/2)}   \right) , \label{doubletY}  \nonumber
\end{eqnarray}
with $\eta(\tau)$ the Dedekind eta-functionis defined by
\begin{align}
	\eta(\tau) = q^{1/24} \prod_{n =1}^\infty (1-q^n)~,
\end{align}
where $q = e^{2 \pi i \tau}$. This doublet modular forms have the following  $q$-expansions:
\begin{align}
	Y^{(2)}_{\bf 2}=\begin{pmatrix}Y_1(\tau)                      \\Y_2(\tau)
	\end{pmatrix}_{\bf 2} =
	\begin{pmatrix}
	\frac{1}{8}+3q+3q^2+12q^3+3q^4+\dots                          \\
	\sqrt{3}q^{1/2}(1+4q+6q^2+8q^3+\dots ) \end{pmatrix}_{\bf 2}.
\end{align}
 The modular forms of weight 4, that is $Y^{(4)}$, can be constructed by using the tensor product of the two doublets $(Y_1(\tau), Y_2(\tau))^T$
\begin{align}
	{\bf 1}~ & : ~~ Y^{(4)}_{\bf 1} = \left(Y_1(\tau)^2+Y_2(\tau)^2\right)_{\bf 1} , \\
	{\bf 2}~ & : ~~ Y^{(4)}_{\bf 2} =
	\begin{pmatrix}
	Y_1(\tau)^2 - Y_2(\tau)^2  \\
	-2Y_1(\tau)Y_2(\tau)
	\end{pmatrix}_{\bf 2} .
\end{align}
The $S_3$ singlet ${\bf 1}'$ modular form of the weight 4 vanishes.


\begin{thebibliography}{99}
\vspace{-1mm}

\bibitem{Feruglio:2017spp}
F.~Feruglio,
doi:10.1142/9789813238053\_0012
[arXiv:1706.08749 [hep-ph]].

\bibitem{Criado:2018thu}
J.~C.~Criado and F.~Feruglio,
SciPost Phys. \textbf{5} (2018) no.5, 042
doi:10.21468/SciPostPhys.5.5.042
[arXiv:1807.01125 [hep-ph]].

\bibitem{deAdelhartToorop:2011re}
R.~de Adelhart Toorop, F.~Feruglio and C.~Hagedorn,
Nucl. Phys. B \textbf{858} (2012), 437-467
doi:10.1016/j.nuclphysb.2012.01.017
[arXiv:1112.1340 [hep-ph]].


\bibitem{S_3}  T. Kobayashi, K. Tanaka, and T.H. Tatsuishi, Phys. Rev. D
98, 016004 (2018); T. Kobayashi, Y. Shimizu, K. Takagi, M. Tanimoto, T. H.
Tatsuishi, and H. Uchida, Phys. Lett. B 794, 114 (2019).
\bibitem{A_4}  T. Kobayashi, N. Omoto, Y. Shimizu, K. Takagi, M.
Tanimoto, and T.H. Tatsuishi, JHEP11(2018)196; H. Okada and M. Tanimoto, Phys. Lett. B 791,54 (2019); P.P. Novichkov, S. T. Petcov, and M. Tanimoto, Phys. Lett.
B 793, 247(2019).
\bibitem{S_4}  J.T. Penedo and S.T. Petcov, Nucl. Phys. B939, 292 (2019);
P.P. Novichkov, J. T. Penedo, S. T. Petcov, and A. V. Titov,
JHEP04 (2019)005.
\bibitem{A_5}  P.P. Novichkov, J. T. Penedo, S. T. Petcov, and A. V. Titov,
JHEP04 (2019) 174; G.J. Ding, S. F. King, and X. G. Liu, Phys. Rev. D 100,
115005 (2019).
\bibitem{1906.10341}
T.~Kobayashi, Y.~Shimizu, K.~Takagi, M.~Tanimoto and T.~H.~Tatsuishi,
PTEP \textbf{2020} (2020) no.5, 053B05
doi:10.1093/ptep/ptaa055
[arXiv:1906.10341 [hep-ph]].

\bibitem{SUGRA}
A.~H.~Chamseddine, R.~L.~Arnowitt and P.~Nath,
Phys.\ Rev.\ Lett.\ {\bf 49}, 970 (1982);
H.~P.~Nilles,
Phys.\ Lett.\ B {\bf 115}, 193 (1982);
L.~E.~Ibanez,
Phys.\ Lett.\ B {\bf 118}, 73 (1982);
R.~Barbieri, S.~Ferrara and C.~A.~Savoy,
Phys.\ Lett.\ B {\bf 119}, 343 (1982);
H.~P.~Nilles, M.~Srednicki and D.~Wyler,
Phys.\ Lett.\ B {\bf 120}, 346 (1983);
J.~R.~Ellis, D.~V.~Nanopoulos and K.~Tamvakis,
Phys.\ Lett.\ B {\bf 121}, 123 (1983);
J.~R.~Ellis, J.~S.~Hagelin, D.~V.~Nanopoulos and K.~Tamvakis,
Phys.\ Lett.\ B {\bf 125}, 275 (1983);
N. Ohta,
Prog.\ Theor.\ Phys.\ 70 (1983) 542;
L.~J.~Hall, J.~D.~Lykken and S.~Weinberg,
Phys.\ Rev.\ D {\bf 27}, 2359 (1983);

  Fei Wang, Kun Wang, Jin Min Yang, Jingya Zhu, JHEP12(2018)041;\\
  Fei Wang, Wenyu Wang, Jin Min Yang, JHEP03(2015)050;\\
  Kun Wang, Fei Wang, Jingya Zhu, Quanlin Jie, Chinese Physics C, 2018, 42(10): 103109.

\bibitem{GMSB}
M.~Dine, W.~Fischler and M.~Srednicki,
Nucl.\ Phys.\ B {\bf 189}, 575 (1981);\\
S.~Dimopoulos and S.~Raby,
Nucl.\ Phys.\ B {\bf 192}, 353 (1981);\\
M.~Dine and W.~Fischler, Phys.\ Lett.\ B {\bf 110}, 227 (1982);\\
M. Dine and A. E. Nelson, Phys. Rev. {\bf D48}, 1277 (1993);\\
M. Dine, A. E. Nelson and Y. Shirman, Phys. Rev. {\bf D51}, 1362 (1995);\\
M. Dine, A. E. Nelson, Y. Nir and Y. Shirman, Phys. Rev. {\bf D53}, 2658 (1996);\\
G. F. Giudice and R. Rattazzi, Phys. Rept. {\bf 322}, 419 (1999).

\bibitem{AMSB}
L.~Randall and R.~Sundrum,
Nucl.\ Phys.\ B {\bf 557}, 79 (1999);
G.~F.~Giudice, M.~A.~Luty, H.~Murayama and R.~Rattazzi,
JHEP {\bf 9812}, 027 (1998).
\bibitem{BNLW}
C.~Balazs, T.~Li, D.~V.~Nanopoulos and F.~Wang,
JHEP \textbf{09} (2010), 003
doi:10.1007/JHEP09(2010)003
[arXiv:1006.5559 [hep-ph]].

\bibitem{flavor:orbifold} F.J. de Anda and S. F. King, J. High Energy Phys. 07 (2018)057;
 T. Kobayashi, Y. Omura, and K. Yoshioka, Phys. Rev. D 78,
115006 (2008); A. Mutter, E. Parr, and P.K. S. Vaudrevange, Nucl. Phys.
B940, 113 (2019).



\bibitem{king} F.J. de Anda, S. F. King, and E. Perdomo, Phys. Rev. D 101,
015028 (2020).

\bibitem{Adulpravitchai:2009id}
A.~Adulpravitchai, A.~Blum and M.~Lindner,
JHEP \textbf{07} (2009), 053
doi:10.1088/1126-6708/2009/07/053
[arXiv:0906.0468 [hep-ph]].

\bibitem{Altarelli} G. Altarelli, F. Feruglio, and Y. Lin, Nucl. Phys. B775, 31
(2007);\\ T. Kobayashi, H. P. Nilles, F. Ploger, S. Raby, and M. Ratz,
Nucl. Phys. B768, 135 (2007);\\
G. Altarelli, F. Feruglio, and C. Hagedorn, JHEP03 (2008) 052.

\bibitem{Chen:2019ewa}
M.~C.~Chen, S.~Ramos-S\'anchez and M.~Ratz,
Phys. Lett. B \textbf{801} (2020), 135153
doi:10.1016/j.physletb.2019.135153
[arXiv:1909.06910 [hep-ph]].


\bibitem{SPheno}
 W. Porod,
 Comput. Phys. Commun. 153 (2003) 275
 [arXiv:hep-ph/0301101];
 W. Porod and F. Staub,
 Comput. Phys. Commun. 183 (2012) 2458
 [arXiv:1104.1573].

\bibitem{SARAH}
  F.~Staub,
  Comput.\ Phys.\ Commun.\  {\bf 185}, 1773 (2014)
  doi:10.1016/j.cpc.2014.02.018
  [arXiv:1309.7223 [hep-ph]];
  F.~Staub,
  Comput.\ Phys.\ Commun.\  {\bf 184}, 1792 (2013)
  doi:10.1016/j.cpc.2013.02.019
  [arXiv:1207.0906 [hep-ph]];
  F.~Staub,
  arXiv:0806.0538 [hep-ph].


\bibitem{micrOMEGA}
  G.~Belanger, F.~Boudjema, A.~Pukhov and A.~Semenov,
  arXiv:1305.0237 [hep-ph];

  G.~Belanger, F.~Boudjema, A.~Pukhov and A.~Semenov,
  arXiv:1005.4133 [hep-ph];

  G.~Belanger, F.~Boudjema, A.~Pukhov and A.~Semenov,
  arXiv:0803.2360 [hep-ph];

  G.~Belanger, F.~Boudjema, A.~Pukhov and A.~Semenov,
  Comput.\ Phys.\ Commun.\  {\bf 176} (2007) 367
  [arXiv:hep-ph/0607059].

\bibitem{HiggsBounds}
  P.~Bechtle, S.~Heinemeyer, O.~Stal, T.~Stefaniak and G.~Weiglein,
  Eur.\ Phys.\ J.\ C {\bf 75}, no. 9, 421 (2015)
  doi:10.1140/epjc/s10052-015-3650-z
  [arXiv:1507.06706 [hep-ph]].

\bibitem{atlas:gluino} M. Aaboud et al. [ATLAS Collaboration], Phys.\ Rev.\ D 97 (2018), no.11, 112001
doi:10.1103/PhysRevD.97.112001
 [arXiv:1712.02332 [hep-ex]];\\
 T. A. Vami [ATLAS and CMS Collaborations], PoS LHCP 2019 (2019) 168 doi:10.22323/1.350.0168
[arXiv:1909.11753 [hep-ex]].

\bibitem{atlas:stop} The ATLAS collaboration [ATLAS Collaboration], ATLAS-CONF-2019-017; A. M. Sirunyan et al. [CMS Collaboration], arXiv:1912.08887 [hep-ex].

\bibitem{LEP} S. Schael {\it et al.} [ALEPH and DELPHI and L3 and OPAL and SLD and LEP Electroweak
Working Group and SLD Electroweak Group and SLD Heavy Flavour Group
Collaborations], Phys. Rept. 427, 257 (2006).

\bibitem{B-physics} V. Khachatryan et al. [CMS and LHCb Collaborations], Nature 522, 68 (2015).
\bibitem{mu-egamma} A.~M.~Baldini {\it et al.} [MEG Collaboration],
  Eur.\ Phys.\ J.\ C {\bf 76}, no. 8, 434 (2016)
  doi:10.1140/epjc/s10052-016-4271-x
  [arXiv:1605.05081 [hep-ex]].

\bibitem{tau-emugamma}
  B.~Aubert {\it et al.} [BaBar Collaboration],
  Phys.\ Rev.\ Lett.\  {\bf 104}, 021802 (2010)
  doi:10.1103/PhysRevLett.104.021802.
  [arXiv:0908.2381 [hep-ex]].

\bibitem{mu-3e}
  U.~Bellgardt {\it et al.} [SINDRUM Collaboration],
  Nucl.\ Phys.\ B {\bf 299}, 1 (1988).
  doi:10.1016/0550-3213(88)90462-2.

\bibitem{tau-3mu3e}
  K.~Hayasaka {\it et al.},
  Phys.\ Lett.\ B {\bf 687}, 139 (2010)
  doi:10.1016/j.physletb.2010.03.037
  [arXiv:1001.3221 [hep-ex]].

\bibitem{Planck}  P. A. R. Ade et al. [Planck Collaboration], Astron. Astrophys. 571, A16 (2014).

\bibitem{WMAP}   J. Dunkley et al. [WMAP Collaboration], Astrophys. J. Suppl. 180, 306 (2009).


\bibitem{LUX}  D.~S.~Akerib {\it et al.},
  arXiv:1608.07648 [astro-ph.CO].
\bibitem{XENON1T}
  E.~Aprile {\it et al.} [XENON Collaboration],
  arXiv:1805.12562 [astro-ph.CO].

\bibitem{PANDAX}  C. Fu {\it et al.}, Phys. Rev. Lett. 118, 071301 (2017)[arXiv:1611.06553].

\bibitem{LZ}B. J. Mount et al., LUX-ZEPLIN (LZ) Technical Design Report, 1703.09144.

\end{thebibliography}
\end{document}